\documentclass[aps,12pt,prl,superscriptaddress,nobibnotes]{revtex4}
\usepackage{amsmath,amssymb,color,graphics,epsfig}
\usepackage[utf8x]{inputenc}

\newcommand{\be}{\begin{equation}}
\newcommand{\ee}{\end{equation}}
\newcommand{\bea}{\setlength\arraycolsep{2pt} \begin{eqnarray}}
\newcommand{\eea}{\end{eqnarray}}
\newcommand{\nn}{\nonumber}

\def\ft#1#2{{\textstyle{\frac{\scriptstyle #1}{\scriptstyle #2} } }}
\def\fft#1#2{{\frac{#1}{#2}}}

\def\0{{\sst{(0)}}}
\def\1{{\sst{(1)}}}
\def\2{{\sst{(2)}}}
\def\3{{\sst{(3)}}}
\def\4{{\sst{(4)}}}
\def\5{{\sst{(5)}}}
\def\6{{\sst{(6)}}}
\def\7{{\sst{(7)}}}
\def\8{{\sst{(8)}}}
\def\sst#1{{\scriptscriptstyle #1}}

\def\a{\alpha}
\def\b{\beta}

\def\d{\delta}

\def\l{\lambda}

\def\m{\mu}
\def\n{\nu}
\def\r{\rho}
\def\s{\sigma}

\def\t{\tau}

\def\o{\omega}
\def\O{\Omega}

\begin{document}
\begin{center}
{\Large {\bf Negative Corrections to Black Hole Entropy from String Theory}}

\vspace{20pt}

{\large Liang Ma, Yi Pang and H. L\"u}

\vspace{10pt}

{\it Center for Joint Quantum Studies and Department of Physics,\\
School of Science, Tianjin University, Tianjin 300350, China }

\vspace{40pt}

\underline{ABSTRACT}
\end{center}
 We report for the first time that in heterotic string compactified on 4-torus or equivalently IIA string compactified on K3, the leading $\a'$ corrections to the rotating black string entropy at fixed conserved charges can be negative. This further implies that the correction to the mass of extremal rotating string is positive, opposite to the standard expectation from the weak gravity conjecture. Our result suggests that the validity of positivity of entropy shift due to higher order operators depends on other factors omitted previously in the effective field theory analysis.

\vfill{\footnotesize  mrhonglu@gmail.com \ \ \ liangma@tju.edu.cn \ \ \ pangyi1@tju.edu.cn
 }

\thispagestyle{empty}
\pagebreak

The last few decades have witnessed the success of effective field theory (EFT) across a broad area of physics. Based on the totalitarian principle that ``Everything not forbidden is compulsory,'' EFT offers a universal framework not only allowing us to decipher experiment data but also paving the way for new predictions without relying on details of the most fundamental theory. Clearly, in applying EFT it is crucial to understand what is actually forbidden by the law of physics. The recent development of swampland program is precisely devoted to answer this question tentatively, concerning consistent couplings to quantum gravity \cite{Palti:2019pca,vanBeest:2021lhn,Harlow:2022gzl}.

In this paper, we focus on the particular constraint stating that corrections to Bekenstein-Hawking entropy of a black hole at fixed mass and charge satisfy \cite{Cheung:2018cwt}
\be
\label{ds}
\Delta S>0
\ee
under the assumptions that for i) the black hole is thermodynamically stable;
ii) Wilson coefficients of higher-dimensional operators are dominated by the tree level contributions from heavy degrees of freedom. Notice that different from the Weak Gravity Conjecture (WGC) constraint \cite{Arkani-Hamed:2006emk}, the inequality \eqref{ds} applies to general stable black holes, not merely the extremal ones. It has been applied to constrain coefficients of higher order operators of Einstein-Maxwell theory utilizing static charged black holes \cite{Kats:2006xp,Cheung:2018cwt}. Subsequent studies have generalized it to other cases by considering variants of Einstein-Maxwell theory or $p$-form theory with additional ingredients \cite{Cano:2019oma, Loges:2019jzs,Cremonini:2019wdk,Ma:2021opb,Ma:2022nwq, Noumi:2022ybv}. In top-down models \cite{Cano:2019oma, Ma:2021opb,Ma:2022nwq}, where the structure of higher order operators are known explicitly, one can indeed find inequality \eqref{ds} holds for static charged black holes and dyonic black strings all the way beyond the extremality.

The derivation of \eqref{ds} utilized an inequality assumption. We denote the light and heavy fields by $\phi$ and $h$ respectively. Given a profile of $(\phi,\, h)$ obeying the field equations of the UV theory, the Euclidean action of the UV theory must satisfy the inequality below 
\be
\label{dI}
I_{UV}[\phi,\,h]<I_{UV}[\phi,\,0]\,.
\ee
At first sight, this inequality appears reasonable, since $(\phi,\, h)$ comprises a stable saddle point while $(\phi,\, 0)$ is not. For ordinary matter whose Euclidean action is
bounded from below and consists of finite number of terms, the inequality above is natural as one can verify that the saddles are indeed the local minimum. However, in a gravitational theory, its full Euclidean action is not bounded from below \cite{Gibbons:1978ac}.  The action is also sensitive to the boundary terms and different choices define different ensembles relevant for the stability analysis. Furthermore if the massive modes arise from a UV theory akin to string theory, the local form of the UV action does not even exist. These considerations suggest there may be string theory examples violating \eqref{ds}. Previous cases that confirmed \eqref{ds} were charged static configurations. Therefore, our focus will be on stationary solutions carrying angular momentum. We are aware that
some work \cite{Cheung:2019cwi,Aalsma:2022knj} have imposed \eqref{ds} on rotating charged black holes to constrain the parameter space of certain EFT. In this paper, we present the first example that violates \eqref{ds} using a well-known stable rotating string solutions in a top-down model from string theory.  Our results thus demonstrate that inequality \eqref{ds} cannot be true in general for the current assumptions.

The specific example we consider is from $T^4$ compactification of heterotic string or equivalently K3 compactification of type IIA string which are $S$ dual to each other. For the purpose of this paper, we can restrict to the bosonic NS-NS sector consisting of the metric $g_{\m\n}$, dilaton $\phi$ and 2-form $B_{\m\n}$. Including leading stringy corrections, the low energy effective of heterotic string compactified on $T^4$ is given by the Bergshoeff-de Roo (BdR) action \cite{Bergshoeff:1989de}
\be
\label{BdR}
{\cal L}_{\rm BdR}= e^{2\phi}\Big[R+4g^{\m\n}\partial_{\m}\phi\partial_{\n}\phi-\ft1{12}G_{\m\n\r}G^{\m\n\r}+
\ft{\a'}2(G^{\m\n\r}\o^L_{\m\n\r}(\widetilde{\o}_-)-\ft12R_{\m\n\r\l}(\widetilde{\o}_-)R^{\m\n\r\l}(\widetilde{\o}_-))\Big],
\ee
where $\a'=\ell_s^2$ is the string length squared and we have defined
\be
G_{\m\n\r}=3\partial_{[\m} C_{\n\r]},\quad \widetilde{\o}_{\m-}{}^{ab}=\o_{\m}{}^{ab}-\ft12G_{\m}{}^{ab},\quad \o_{\m\n\r}^L(\widetilde{\o}_-)={\rm tr}(\widetilde{\o}_{[\m-}\partial_{\n}\widetilde{\o}_{\r-]}+\ft23\widetilde{\o}_{[\m-}\widetilde{\o}_{\n-}\widetilde{\o}_{\r-]})\ ,
\ee
in which $\o_\m{}^{ab}$ is the ordinary spin connection.
On the other hand, the low energy effective action of K3 compactification of type IIA string theory with leading stringy corrections on K3 takes the form \cite{Liu:2013dna, Liu:2019ses}
\be
\label{IIA}
 I_{IIA}=\frac{1}{16\pi G_6}\int d^6x (\mathcal{L}_{2\partial}+
\ft{\a'}{16}\mathcal{L}_{4\partial})\,,
 \ee
in which the 2- and 4-derivative Lagrangians are given by
\bea
\label{2dL}
\mathcal{L}_{2\partial}&=&\sqrt{-g}e^{-2\phi}\Big(R+4g^{\m\n} \partial_\m\phi\partial_\n \phi-\ft{1}{12}H_{\mu\nu\rho}H^{\mu\nu\rho}\Big),\quad H_{\m\n\r}=3\partial_{[\m}B_{\n\r]}\,,
\nn\\
\label{hdL2}
\mathcal{L}_{4\partial}&=&\sqrt{-g}\left[2R_{\mu\nu\rho\sigma}R^{\mu\nu\rho\sigma}-
4R_{\mu\nu}R^{\mu\nu}+R^2+\ft16RH^2
-R^{\mu\nu}H^2_{\mu\nu}-\ft{7}{24}H_4\right.\cr
&&\left.+\ft{1}{144}(H^2)^2-\ft{3}{8}(H^2_{\mu\nu})^2
+\ft{1}{3}\epsilon^{\mu\nu\lambda\rho\sigma\tau}H_{\mu\nu\l}(\o^L_{\rho\sigma \tau}(\o_+)+\o^L_{\rho\sigma \tau}(\o_-))\right],
\eea
where the 4-derivative action is a combination of the  supersymmetric Riemann squared term \cite{Bergshoeff:1986wc} and the supersymmetric Gauss-Bonnet invariant \cite{Novak:2017wqc,Butter:2018wss} with equal coefficients. In equations above the definitions below are used
\bea
H_{\m\n}^2&:=&H_{\m\r\s}H_{\n}{}^{\r\s},\quad H^2:=H_{\m\n\r}H^{\m\n\r},\quad H_4:=H_{\m\n\s}H_{\r\l}{}^{\s}H^{\m\r\d}H^{\n\l}{}_{\d}\ ,
\nn\\
{\o}_{\m\pm}{}^{ab}&=&\o_{\m}{}^{ab}\pm\ft12H_{\m}{}^{ab},\quad \o_{\m\n\r}^L({\o}_\pm)={\rm tr}({\o}_{[\m\pm}\partial_{\n}{\o}_{\r\pm]}+\ft23{\o}_{[\m\pm}{\o}_{\n\pm}{\o}_{\r\pm]})\,.
\eea
It is long expected that the two Lagrangians in \eqref{BdR} and \eqref{IIA} are related to each other by $S$-duality \cite{Hull:1994ys}. By explicit computation the recent work \cite{Chang:2022urm} confirmed that up to terms removable by field redefinitions, the heterotic string action \eqref{BdR}
can be derived from the IIA string action \eqref{IIA} via dualizing $H_{\m\n\r}$ to $G_{\m\n\r}$ and rescaling the metric $g_{\m\n}\rightarrow e^{2\phi}g_{\m\n}$. Thus onshell we must have
\be
\label{dualBdR}
{\cal L}_{\rm BdR}={\cal L}_{IIA}+\frac1{12}\epsilon^{\m\n\r\s\t\l}H_{\m\n\r}\partial_{\s}C_{\t\l},\quad g^{IIA}_{\m\n}=e^{2\phi}g^h_{\m\n}\ .
\ee
According to \cite{Reall:2019sah}, when expressed as a function of conserved charges, the leading higher derivative corrections to the entropy can be obtained by simply evaluating the 4-derivative action on  the solution of the 2-derivative theory. Thus from \eqref{dualBdR}, it is evident that for black strings in heterotic and IIA related by dualization of 3-form field strength, leading higher derivative corrections to the entropy are identical module the fact that the electric and magnetic charges carried by the 3-form field strength are interchanged. In fact the equality \eqref{dualBdR} plays a crucial role in computing the 4-derivative corrections to the entropy of heterotic black string. One notices that the single Lorentz Chern-Simons (CS) term in the BdR action \eqref{BdR} does not transform covariantly under local Lorentz transformation. Thus when evaluating the Lorentz CS term on the uncorrected solution, the result appears to be  non-unique depending on the choice of frame. One may replace the Lorentz CS term by its counterpart made from the Levi-Civita connection. However, in this case, a direct computation shows that the CS term contains a certain component ill-defined on the $S^3$ direction of the black string solution \cite{Ma:2021opb}. In contrast, the 4-derivative terms in the IIA action \eqref{IIA} does not suffer from these problems. The ambiguities or singularities from the pair of
Lorentz CS term cancel. Also the general coordinate invariance can be restored by recasting the $H_{(3)}\wedge(\o^L(\o_+)+\o^L(\o_-))$ term as $-H_{(3)}\wedge({\rm tr}(R(\o_+)\wedge R(\o_+))+{\rm tr}(R(\o_-)\wedge R(\o_-)))$. We have verified that these two expressions give the same result although they differ by  a total derivative term. Thus utilizing \eqref{dualBdR} we can compute the leading stringy corrections to the black string entropy in both heterotic and type IIA string theory in one go. For convenience, we choose to work in the IIA frame.

The 2-derivative theory in \eqref{IIA} admits the rotating 3-charge black string solution \cite{Cvetic:1998xh} obtained from lifting the general rotating black holes in five-dimensional ungauged STU model \cite{Cvetic:1996xz,Wu:2011gq,Chow:2016hdg}. For simplicity, we consider here the special case where the string carries two equal angular momenta. To present the solution, we  define $s_I=\sinh\delta_I$, $c_I=\cosh\delta_I$, $c_{123}=c_1c_2c_3$, $s_{123}=s_1s_2s_3$ and 1-forms $K_I,\,I=1,2,3$, as
\be
K_I=s_Ic_Idt-a(c_{123}\frac{s_I}{c_I}-s_{123}\frac{c_I}{s_I})(\sin^2\theta d\phi_1+\cos^2\theta d\phi_2)\,.
\ee
The six-dimensional metric and dilaton of the equally rotating string can be put in the form
\be
\label{CL}
ds_6^2=ds^2_5+\frac{X_1}{X_3}(dz+A_3)^2\,,\qquad e^{-2\phi}=\frac{H_1}{H_2}\,,
\ee
where denoting $\rho^2=r^2+a^2$, various quantities above are
\bea
H_I&=&1+\frac{2ms^2_I}{\rho^2}\,,\qquad X_I=\frac{(H_1H_2H_3)^{1/3}}{H_I}\,,\qquad A_3=\frac{2m}{H_3\rho^2}K_3\,,\nn\\
ds_5^2&=&H_2\Big(-dt^2+\frac{\rho^2}{\Delta_r}dr^2+\rho^2d\Omega^2_3
+\sum_{I=1}^3\frac{(1-1/H_I)K^2_I}{\prod_{J\neq I}(s^2_I-s^2_J)}\Big),\nn\\
d\Omega^2_3&=&d\theta^2+\sin^2\theta d\phi^2_1+\cos^2\theta d\phi^2_2\,,\qquad \Delta_r=\frac{\rho^4}{r^2}-2m\,.
\eea
The 2-form potential is given by
\bea
B_{(2)}&=&\frac{m}{\rho^2}\Big((\frac{1}{H_1}+\frac{1}{H_3})
-\frac{2m}{\rho^2}\frac{1}{H_1H_3}(s_1^2-s_3^2)
\Big) dt\wedge K_2\cr
&&+\frac{2m}{H_1\rho^2}K_1\wedge dz+2mc_2s_2\cos^2\theta d\phi_1\wedge d\phi_2\ .
\eea
This solution carries energy $E$, linear momentum $P_z$ along $z$-direction, electric and magentic charges
$(Q_e, Q_m)$, and two equal angular momenta $J_1=J_2$ in the 4-plane transverse to the string. Their expressions are presented in \eqref{etsj} with $a=b$.

The event horizon of the black string solution is located at $r_h=\frac{1}{\sqrt{2}}(\sqrt{m}+\sqrt{m-2a^2})$, with temperature given by
\be
T_0=\frac{(r_h^2-a^2) (r_h^2+a^2)}{4 \pi  m r_h (r_h^2c_{123}+a^2 s_{123}  )}\ .
\ee
Henceforth we use subscript ``0" to label quantities without $\a'$ corrections.
The entropy density of the string \eqref{CL} can be computed using Wald formula \cite{Wald:1993nt} yielding
\be
S_0=\frac{\pi^2}{r_h G_6}m(r^2_hc_{123}+a^2s_{123})\,.
\ee

Turning on the $\a'$ corrections to the effective action, the solution will be modified.  Accordingly, the entropy receives corrections proportional to $\a'$. The solution carries a set of conserved charges denoted as $\vec{{\cal Q}}$, in terms of which the $\a'$ correction to the entropy density can be parameterized as
\be
S(\vec{{\cal Q}})=S_0(\vec{{\cal Q}})+\Delta S(\vec{{\cal Q}})\,.
\ee
The leading $\a'$
corrections to entropy density can be readily obtained by evaluating
the $\a'$ part of the effective action \eqref{IIA} on the uncorrected solution \eqref{CL} \cite{Reall:2019sah}. The result is simple
\be
\label{Ihd}
\Delta S(\vec{{\cal Q}})=\frac{\pi\b_0}{32 G_6}\frac{(9r_h^4-14r_h^2a^2+a^4)c_1^2-(r_h^4-14r_h^2a^2+9a^4)s_1^2}{(r_h^2c_1^2+a^2s_1^2)^2}
\alpha'\,,\ee
where $\beta_0=T_0^{-1}$, and parameters $r_h,\, a,\, \d_1$ should be thought of as implicit functions of the conserved charges. From the expression above, one can see that when rotating parameter $a=0$, $\Delta S(\vec{{\cal Q}})$ is always positive, reproducing the previous results \cite{Ma:2021opb,Ma:2022nwq}. On the other hand, when $a$ is non-zero, $\Delta S$ has no definite sign. In particular, $\Delta S(\vec{\cal Q})=0$
when
\bea
\fft{a}{r_h}&=& \sqrt{\frac{7\pm2 \sqrt{10 +16 c_1^2 s_1^2}}{1-8 s_1^2}},\qquad 1-8s^2_1>0\,,
\nn\\
 \fft{a}{r_h}&=& \sqrt{\frac{2 \sqrt{10 +16 c_1^2 s_1^2}-7}{8 s_1^2-1}},\qquad 1-8s^2_1<0\,.
\label{alpha}
\eea
In the argument leading to the positive entropy shift, a necessary condition is the stability of the uncorrected solution. To find a counterexample, we must show that the stability and the negative entropy shift can occur simultaneously in a certain parameter regime.

Notice that fields in the effective action carry neither electric nor magnetic charges associated with the 2-form $B_{\m\n}$; therefore, perturbations of fields will not generate fluctuations of those two charges. This point of view was first suggested by \cite{Reall:2001ag} to show that most of the black strings and branes in string and M theory are stable near extremality \footnote{In the original paper \cite{Cheung:2018cwt} in order to show WGC constraint can be derived from the inequality \eqref{ds}, the electric charge must be fixed in the ensemble; otherwise, there is no stable static charged black holes on which \eqref{ds} can be applied.}. In the ensemble with fixed electric and magnetic charges, only
the energy, linear momentum and angular momentum are allowed to vary. The thermodynamic ensemble under consideration is also consistent with the variation
principle of the action with appropriate boundary terms added. Under generic perturbations about the equally rotating string solution \eqref{CL}, stability of the system requires the Hessian of the energy
\be
\frac{\partial^2 E}{\partial x^\mu\partial x^\nu}\,,\qquad x^\mu=(S,\, P_z,\, J_1,\,J_2)\,,
\ee
be positive definite \cite{Monteiro:2009tc}. The Hessian involves second derivatives with respect to 4 variables because perturbations can carry unequal angular momenta in the plane transverse to the string. In the following, we investigate the possitivity of the Hessian using the leading order thermodynamic quantities since the corrections due to higher derivative interactions are suppressed by dimensionless parameter $\ell_s^2/r_h^2$ which is much smaller than the leading term.
 Thermodynamic quantities of an uncorrected general rotating 3-charge string solution are \cite{Wu:2011gq} (See also \cite{Cvetic:2004ny,Chong:2005hr}.)
\bea
\label{etsj}
E_0&=&\frac{\pi}{4 G_6}m(c_1^2+s_1^2+c_2^2+s_2^2+c_3^2+s_3^2)\,,\qquad m=\frac{\left(r_h^2+a^2\right) \left(r_h^2+b^2\right)}{2 r_h^2}\ ,\cr
%%%
T_0&=&\frac{\left(r_h^2-a b\right) \left(r_h^2+a b\right)}{4 \pi  m r_h \left(r_h^2c_{123}+a b s_{123} \right)}\,,\qquad S_0=\frac{\pi^2}{r_h G_6}m(r^2_hc_{123}+abs_{123})\,,\cr
%%%
\Omega_{1,0}&=&\frac{r_h^2a}{(r^2_h+a^2)(r^2_hc_{123}+abs_{123})}\,,\qquad J_{1,0}=\frac{\pi}{2 G_6}m(ac_{123}-bs_{123})\,,\cr
\Omega_{2,0}&=&\frac{r_h^2b}{(r^2_h+b^2)(r^2_hc_{123}+abs_{123})}\,,\qquad J_{2,0}=\frac{\pi}{2 G_6}m(bc_{123}-as_{123})\,,\cr
\Phi_{I,0}&=&\frac{r^2_hc_{123}\frac{s_I}{c_I}+abs_{123}\frac{c_I}{s_I}}{r^2_h c_{123}+abs_{123}} \,,\qquad Q_{I,0}=\frac{\pi}{2 G_6}mc_Is_I\,,\quad I=1, 2, 3,
%\Phi_{0m}&=&-\frac{r^2_hc_{123}\frac{s_2}{c_2}+abs_{123}\frac{c_2}{s_2}}{r^2_hc_{123}+abs_{123}}
%\,,\qquad Q_{0 m}=-\frac{\pi}{2 G_6}mc_2s_2\,,\cr
%V_{0z}&=&\frac{r^2_hc_{123}\frac{s_3}{c_3}+abs_{123}\frac{c_3}{s_3}}{r^2_hc_{123}+abs_{123}}
%\,, \qquad P_{0z}=\frac{\pi}{2 G_6}mc_3s_3\,.
\eea
where $Q_{I,0}$ and $\Phi_{I,0}$ stand for the charges $Q_{e,0}$, $Q_{m,0}$, $P_{z,0}$ and their thermodynamic conjugate potentials $\Phi_{e,0}$, $\Phi_{m,0}$ and $V_{z,0}$ respectively.
Before computing the Hessian of energy with $Q_{e,0}$ and $Q_{m,0}$ fixed, we adopt a procedure simplifying the calculation. Since the positivity of the Hessian is invariant under change of variable, we switch from the original variables $x^\m=(S_0,\, P_{z,0},\, J_{1,0},\,J_{2,0})$ to new variables $y^\m=(a,\,b,\,r_h,\, e^{\d_3})$. Since the electric and magnetic charges are fixed, $e^{\d_1}$ and $e^{\d_2}$ are dependent on other variables. Using the fact that \cite{Liu:2010sz}
\be
\label{he}
\frac{\partial^2 E}{\partial x^\m\partial x^\n}dx^\m dx^\n=dTdS+d\O_i dJ_i+dV_z dP_z+d\Phi_{e}dQ_{e}+d\Phi_{m}dQ_{m}\,,
\ee
the Hessian matrix in the basis of new variables is readily obtained by plugging the thermodynamic quantities into \eqref{he} and using the non-holonomic constraints
\bea
d e^{\delta_1}&&=-\frac{2e^{\delta _1} c_1  s_1 \left(  (r_h^4-a^2 b^2 )d r_h+b r_h  (a^2+r_h^2 )d b+a  r_h  (b^2+r_h^2 )da\right)}
{ r_h  (r_h^2+a^2 ) (r_h^2+b^2 ) (c_1^2+s_1^2 )}\,,\cr
d e^{\delta_2}&&=-\frac{2e^{\delta _2} c_2  s_2 \left(  (r_h^4-a^2 b^2 )d r_h+b r_h  (a^2+r_h^2 )d b+a  r_h  (b^2+r_h^2 )da\right)}
{ r_h  (r_h^2+a^2 ) (r_h^2+b^2 ) (c_2^2+s_2^2 )}\,,
\eea
implied by $dQ_{e,0}=0$ and $dQ_{m,0}=0$.
We then calculate the principal minors of the Hessian matrix denoted by
$\Xi_i,\, i=1,\cdots,4$ corresponding to the equally rotating string solution with $a=b$.

In search of the parameter regime where $\Xi_i>0$ and $\Delta S<0$, based on \eqref{alpha} we introduce a dimensionless parameter $x$ away from the $\Delta S=0$ point. Specifically we consider 
\be
\frac{a}{r_h}=\sqrt{\frac{2 \sqrt{10 +16 c_1^2 s_1^2}-7}{8 s_1^2-1}} + x\,.
\ee
There are two other ways of introducing a dimensionless parameter away from $\Delta S=0$ since this equation admits three branches of solution for $a/r_h$ \eqref{alpha}. However, to show that there exist parameter regions where $\Delta S$ can be negative while the solution is thermodynamically stable, one choice suffices for this purpose.

\begin{figure}[hbtp]
  \centering
  \includegraphics[width=0.48\textwidth]{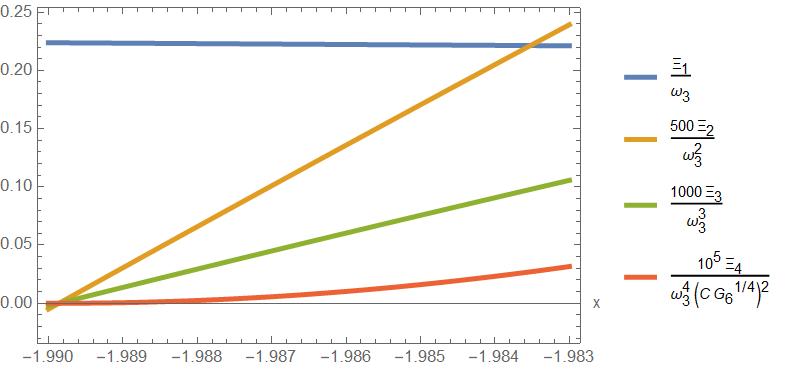}\ \
  \includegraphics[width=0.48\textwidth]{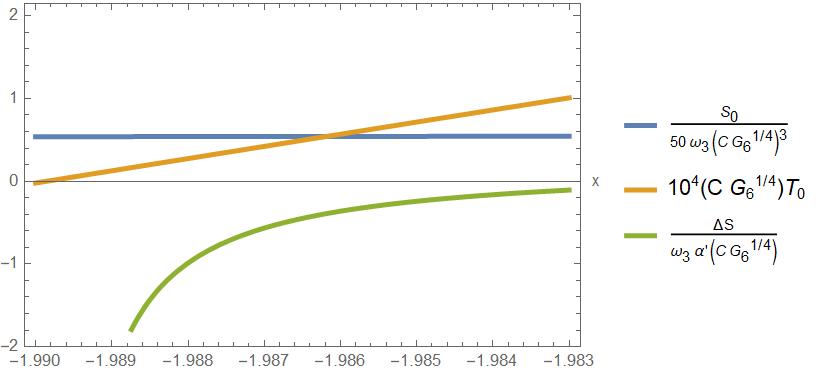}\
  \includegraphics[width=0.49\textwidth]{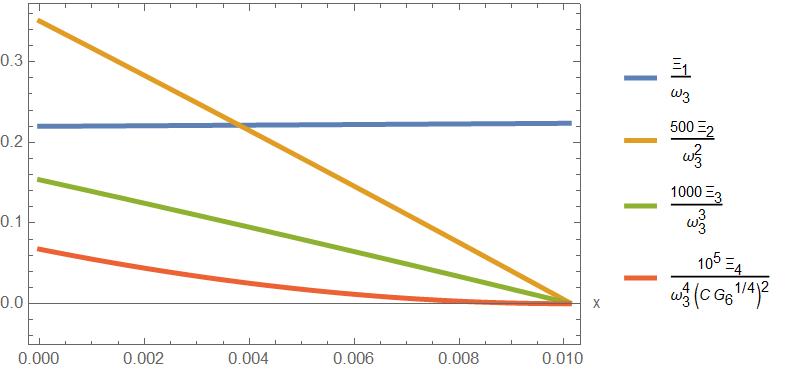}\
  \includegraphics[width=0.49\textwidth]{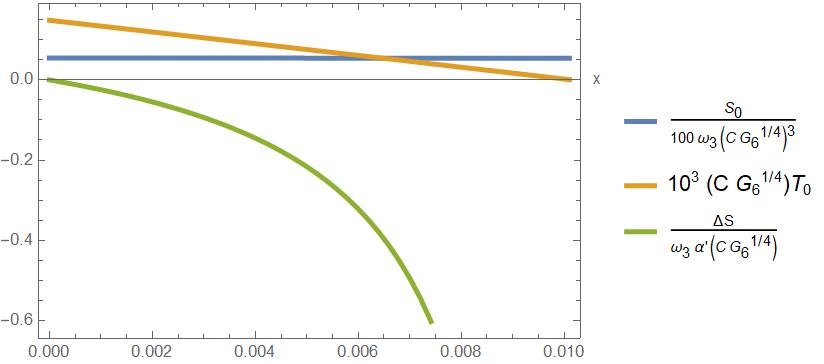}\
 \caption{\small These plots exhibit the behaviors of the four principal minors, the entropy shift, uncorrected temperature and entropy. The top and bottom two figures have $x\in[-1.9898,-1.9797]$ and $x\in[0,0.0101]$ respectively. We have substituted $\omega_3$ for $2\pi^2$ which is the area of a unit 3-sphere. }
\label{fixQeQm1}
\end{figure}

 After some trial and error, we do find parameters where the stable rotating string can have negative entropy shift generated by higher derivative string corrections. Such parameter regions are actually ample. For instance, with the choice of parameters
\bea
\quad e^{\delta_1}=7,\qquad e^{\delta_2}=\frac{1}{10},\qquad e^{\delta_3}=\frac{6}{5},\qquad m=C^2\sqrt{G_6}\ ,
\eea
$Q_{e,0}$, $Q_{m,0}$ and $P_{z,0}$ now take fixed values. We do not specify $C$, as it factorizes out
from principal minors and thermodynamic quantities. The one free parameter left is $x$. We observe that when $x \in [-1.9898,\,-1.9797]\cup [0,\,0.0101]$, the 4 principal minors are positive but $\Delta S$ is negative. We have also checked that temperature and entropy stays positive in the same parameter region. The results are presented in Fig.~\ref{fixQeQm1}
%%%

%\begin{figure}[hbtp]
%  \centering
%  \includegraphics[width=0.48\textwidth]{fixQeQm008PM122.jpg}\ \
%  \includegraphics[width=0.48\textwidth]{fixQeQm008PM342.jpg}\
%  \includegraphics[width=0.49\textwidth]{fixQeQm008deltaS2.jpg}\
%  \includegraphics[width=0.49\textwidth]{fixQeQm008T0S02.jpg}\
% \caption{This figure exhibits the behaviors of 4 principal minors, the entropy shift, %uncorrected temperature and entropy within $x\in[0,0.0101]$. We have substituted $\omega_3$ %for $2\pi^2$ which is the area of a unit 3-sphere. }\label{fixQeQm2}
%\end{figure}
%%%
Finally, we discuss the limit Bogomol'nyi-Prasad-Sommerfield (BPS) limit of the string solution. The BPS limit is attained by taking $r_h\rightarrow 0$ with $a=\tilde a r_h,  b=\tilde b r_h$, while keeping the quantities $q_I=2m s_I^2$ finite. ($s_I\rightarrow -\infty$ to be precise.) The corresponding charges are
\be
Q_e=\frac{\pi}{G_6}q_1,\qquad Q_m=\frac{\pi}{G_6}q_2,\qquad P_z=\frac{\pi}{G_6}q_3,\qquad J_1=J_2=\frac{\pi}{4G_6}j\,,
\ee
where $j= 8(\tilde a+\tilde b ) \sqrt{q_1q_2q_3}/\sqrt{(1+\tilde a ^2)(1+\tilde b^2)}$. (We introduce a universal minus in the definition of $(Q_e,Q_m, P_z)$.) The limit above results in the energy saturating the BPS bound
\be
E=Q_e+Q_m+P_z\,,
\ee
and the total entropy
\be
S_{\mathrm{BPS}}=\frac{\pi^2}{2 G_6}
\sqrt{q_1 q_2 q_3-j^2}\left(1+\frac{\alpha'}{q_1}\right),
\ee
which recovers the previous result of \cite{Ma:2022nwq} when rotation is switched off. We see that the entropy shift in the BPS is positive.

In this paper, we reported for the first time that in $T^4$ compactification of heterotic string or K3 compactification of type IIA string, the leading $\a'$ corrections to the rotating black string entropy at fixed conserved charges can be negative. Angular momentum plays an indispensable role in our result. Had it been zero, the entropy shift would always be positive. In the heterotic case, the 4-derivative corrections appear at tree level thus providing a concrete counterexample to the proposal of \cite{Cheung:2018cwt}. The results of \cite{Cheung:2018cwt,Goon:2019faz,Crisford:2017gsb} suggest that the weak gravity conjecture, positivity of entropy shift and cosmic censorship are connected to each other. Usually when one of them is obeyed, the others are 
also satisfied. Thus the fact that angular momentum plays an important role in our findings of negative entropy shift seems to resonate with the violation of the cosmic censorship observed in the spinning black holes \cite{Figueras:2017zwa}.

Our results imply that $\Delta S>0$ may not be valid in general and its validity requires further clarifications.  Applying formulae given in \cite{Goon:2019faz}, this result also implies the correction to the mass of extremal rotating string is positive, as in \cite{Ma:2020xwi}, opposite to the expectation from WGC, which in the static case, simply follows from $\Delta S>0$. Clearly this indicates the inequality is still quite powerful when used properly. Thus it should be very useful to understand it better by testing various black holes solutions in other top-down models, before adopting it as a general criteria constraining effective field theories.

\section*{Acknowledgement}

L.M. thanks Dr. Zhan-Feng Mai for graphic advice. The work is supported in part by the National Natural Science Foundation of China (NSFC) grants No.~11875200, No.~11935009 and No.~12175164. This work of Y.P. is also supported by the National Key Research and Development Program under grant No. 2022YFE0134300.

\end{document}